# Modular2Simple: A Tool for Modular Scenario Creation Based on the OpenSCENARIO Format

Nikolai Khriapov[1*], Mohamed Taha Drif[1], Renjue Li[2] and Cas Widdershoven[2]

[1] Nanjing University of Aeronautics and Astronautics, Nanjing 210016, People's Republic of China
[2] Chinese Academy of Sciences, Beijing 100190, People's Republic of China
nikolai.khriapov@nuaa.edu.cn

**Abstract.** The rapid advancement of autonomous driving systems (ADS) has introduced significant challenges, particularly in the creation of realistic and complex scenarios for testing and validation. This paper introduces Modular2Simple, a tool designed to address these challenges by simplifying and enhancing the process of creating complex ADS scenarios. Modular2Simple seamlessly integrates with the CARLA simulator and is applicable to any software that supports the OpenSCENARIO format. By leveraging existing simple scenarios in the OpenSCENARIO format, the tool enables developers to create easily customizable modular scenarios through the combination of multiple simple or modular scenarios, significantly simplifying the scenario creation process while maintaining flexibility in scenario design. This approach not only facilitates the development of complex scenarios, reducing both development time and effort, but also promotes scenario reuse and customization, which leads to a significant reduction in code complexity and enhanced efficiency in scenario design and testing compared to traditional scenario development methods.

## 1 Introduction

Autonomous vehicles (AV) are on the brink of transforming our daily lives and revolutionizing transportation. However, achieving fully autonomous driving requires rigorous testing, verification, and validation. Simulations play a fundamental role, as they provide a controlled and safe environment for testing various aspects of autonomous systems, from perception and decision-making to vehicle control.

In the field of autonomous driving systems, the creation of complex traffic scenarios holds a pivotal role in enhancing the safety and efficiency of self-driving vehicles. This paper introduces Modular2Simple, a tool designed to aid in the process of creating complex scenarios for ADS simulations.

At its core, Modular2Simple empowers developers to construct modular scenarios by combining multiple simple scenarios or pre-existing modular scenarios. This approach simplifies the development of highly complex and realistic scenarios crucial for comprehensive ADS testing. It encourages not just the creation of scenarios but also their reuse and customization, a practice that fosters efficiency and consistency in testing procedures.

This paper provides a comprehensive overview of Modular2Simple, showcasing its functionalities, applications, and the advantages it offers to ADS research and development. By leveraging data from simple scenario files and ensuring seamless compatibility with CARLA's ScenarioRunner module, Modular2Simple demonstrates itself as a valuable tool for researchers and developers in the ADS field.

While the primary focus is on the CARLA simulator [1], a well-established platform in this domain, it's important to emphasize that Modular2Simple is applicable to any software supporting the OpenSCENARIO format [2].

Modular2Simple is an open-source tool, with both its code and documentation openly available on GitHub [3].

The remainder of this paper is structured as follows: Section II gives an overview of related work, Section III includes the framework description, Section IV outlines case studies, and Section V concludes this paper with a short summary and outlook.

## 2 Related Works

The focus on virtual scenario-based testing for ADS has grown, especially due to regulatory demands requiring extensive testing distances for vehicle acceptance [4]. Various tools and methodologies have been developed to facilitate scenario creation and simulation in the ADS domain.

One prevalent approach involves scenario generation method based on scenario description languages (SDL) or modeling languages, which describe scenario elements and their relationships for creating dynamic scenarios. OpenSCENARIO is a widely used format, alongside other SDLs for specific purposes. For instance, GeoScenario, created in the Waterloo Intelligent Systems Engineering Lab, supports diverse traffic elements and is XML-based, simple, extensible, and built on the Open Street Map (OSM) standard for driving simulations [5]. Similarly, stiEF enhances clarity and precision, guiding users through scenario creation and ensuring completeness, avoiding ambiguities [6]. SceML, from FZI Research Center for Information Technology, uses a graph-based approach. It encodes scenarios into machine-readable structures aligned with UML class diagram rules, facilitating scenario creation and management through graphical modeling, promoting consistency and interoperability in ADS testing [7].

Another significant approach is data-driven scenario creation, leveraging real-world data to construct scenarios. This methodology relies on recorded, concrete, or logical scenarios from various sources, including AVs accident reports. Utilizing such data allows for generating critical scenarios to test algorithmic improvements by learning from past mistakes. For example, Goss et al. [8] used accident reports to generate scenarios to improve AV safety through simulation, emphasizing the importance of accident data. Countries have invested in comprehensive accident data collection, leading to the establishment of accident databases like CIDAS [9] in China and GES [10] in the U.S. Furthermore, methodologies and frameworks have been developed to evaluate AV functions and calibrate traffic flow models using measured data from official highway test sections. Nalic et al. [11] presented a methodology for evaluating a Highway Chauffeur function, highlighting the integration of measured data into scenario evaluation processes. Companies like Waymo use actual driving data to create parameterized situations for extensive testing, running up to 8 million miles per day using 25,000 virtual vehicles [12].

## 3 Framework Description

Modular2Simple empowers developers to construct modular scenarios by combining multiple simple scenarios or pre-existing modular scenarios. Constructing scenarios in a modular fashion allows to simplify the development of highly complex scenarios, and efficiently build and reuse scenarios, thereby streamlining the testing process.

Modular2Simple is implemented in Java and provides a command-line interface for scenario conversion and the management of modular scenario files. Since the output of Modular2Simple is in the OpenSCENARIO format, it is compatible with any software that supports this format. In particular, this includes the CARLA simulator.

Its functionality encompasses two primary operations, as depicted in Fig. 1:

1. Packaging a set of simple or modular scenarios into a single modular scenario:
*java Modular2Simple -cxm <input_xosc_and_mosc_files> <output_mosc_file>*,
2. Converting the resulting modular scenario into a single-file scenario:
*java Modular2Simple -cmx <input_mosc_file> <output_xosc_file>*.

Modular2Simple uses the custom ". mosc" extension for packaging multiple simple or modular scenario files within a single modular scenario, containing all necessary scenarios and a main file referencing these scenarios. The ".mosc" package can include other ".mosc" packages alongside simple scenarios. Scenarios from other directories can be included by specifying the "SCENARIO_PATH" environment variable.

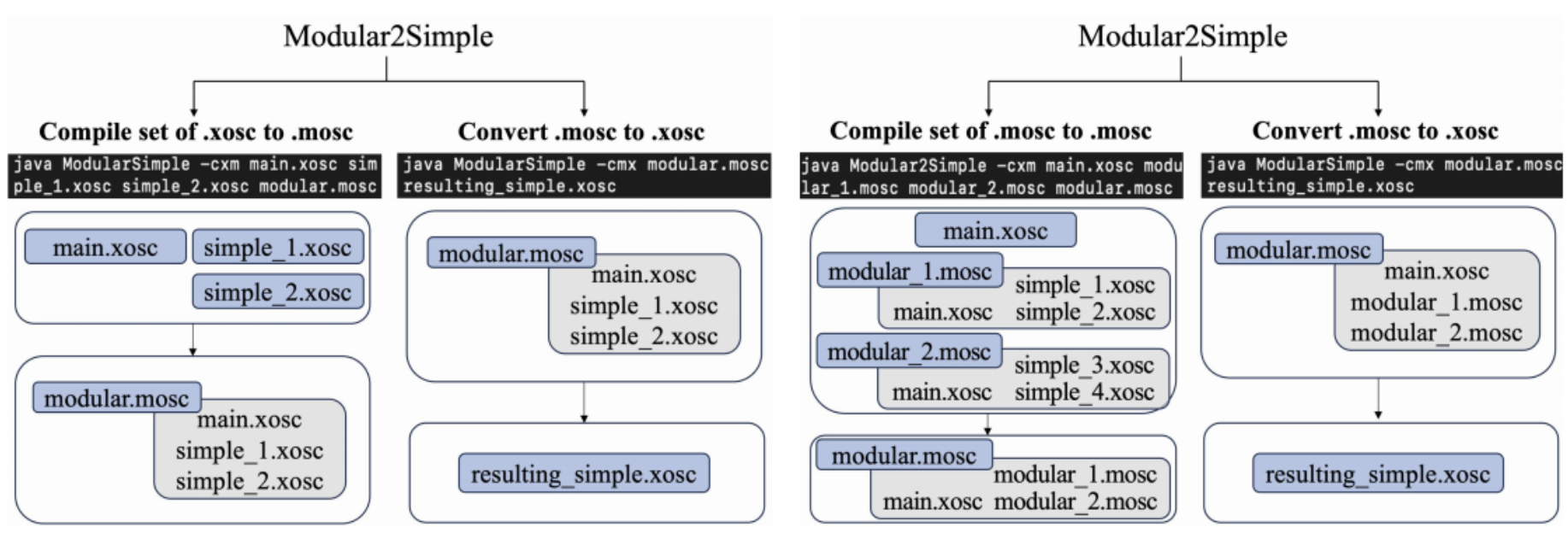

**Fig. 1.** Modular2Simple functionalities overview.

The "main.xosc" file follows the standard OpenSCENARIO format with one critical distinction: instead of directly specifying maneuvers, it references required simple or modular scenario files through a custom ScenarioReference tag, which includes keys and values of attributes that require customization within these scenarios:

*<ScenarioReference scenarioFileName="scenario.xosc">*
*<ManeuverGroupReference maneuverGroupName="name">*
*<ParameterReference key="attribute1" value="value1" />*
*</ManeuverGroupReference>*
*</ScenarioReference>*

Simple scenario files follow the standard OpenSCENARIO format, differing only by incorporating a “parameterAttributes” attribute in the ManeuverGroup tag to specify attributes customizable from the main file. If any attributes are not defined in the main file, default values are used, encouraging the reuse of customized simple scenarios:

*<Condition parameterAttributes=“key1” key1=“defaultValue1”>…</Condition>*

## 4 Case Study

The simulations were conducted using the CARLA simulator (version 0.9.15) on a machine equipped with an Intel i7-10750H CPU, 8GB RAM, and an NVIDIA GTX 1650 GPU to ensure the smooth execution of high-resolution visualizations and complex scenario simulations.

The first case study illustrates a collision between two vehicles at an intersection, leading to the arrival of a police vehicle and an ambulance, after which all vehicles depart, as shown in Fig. 2. The files for this case study can be found on GitHub [3].

The structure of the example modular scenario is shown in Fig. 2. The scenario framework is composed of a single modular scenario that references two smaller modular scenarios. Each smaller scenario references two simple scenarios: one for initiating the collision, one for the arrival of emergency vehicles, and one for the departure of each vehicle. The modular scenario “modular_1” outlines the behavior of a vehicle involved in the accident and departing from the accident scene. This scenario is utilized by both vehicles involved in the collision. The modular scenario “modular_2” outlines the behavior of a vehicle navigating to and departing from the accident site. This scenario is shared by both the police vehicle and the ambulance. Also, since the scenario “simple_3” is shared between “modular_1” and “modular_2”, it could be saved outside this mosc package in a folder defined in SCENARIO_PATH.

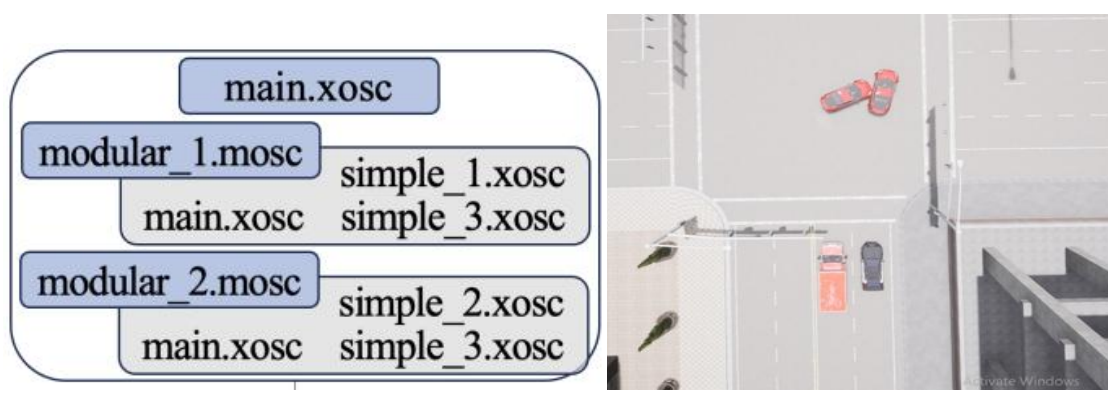

**Fig. 2.** Structure and visual representation of the example modular scenario.

To showcase the effectiveness of modularity, we generated a small scenario library consisting of 31 scenarios featuring different combinations of actors such as pedestrians crossing the road, cars crossing the junction ahead, and cars coming up from behind. The behavior of these actors is controlled by just 2 simple scenarios reused across all modular files. As a result, the modular scenario library just contains 10,811 lines of code, compared to 21,243 lines in a traditional scenario library, achieving a 49% reduction in codebase size. The detailed breakdown of lines of code for individual scenarios, along with the corresponding code reduction percentages, is presented in Table I. The table presents a detailed comparison of the number of lines of code (LoC) for each scenario created using the modular approach versus the traditional

OpenSCENARIO format. The “Code Reduction (%)” column reflects the percentage reduction in code size achieved by using the modular approach. The total lines of code accounted for in the calculation include those from two simple scenarios reused across the modular scenarios, with respective line counts of 247 and 276. Note that code reduction percentage can be improved even further by more extensive reuse of behavior and reuse of more complex behavior. The files for this case study can be found on GitHub [3].

**Table 1.** Lines of Code Comparison for Individual Scenarios.

| Scenario | LoC in Modular2Simple Format | LoC in OpenSCENARIO Format | Code Reduction (%) |
|---|---|---|---|
| 1 | 200 | 358 | 44.1 |
| 2 | 274 | 538 | 49.1 |
| 3 | 205 | 417 | 50.8 |
| 4 | 290 | 608 | 52.3 |
| 5 | 364 | 788 | 53.8 |
| 6 | 278 | 596 | 53.4 |
| 7 | 200 | 358 | 44.1 |
| 8 | 286 | 550 | 48.0 |
| 9 | 360 | 730 | 50.7 |
| 10 | 290 | 608 | 52.3 |
| 11 | 376 | 800 | 53.0 |
| 12 | 450 | 980 | 54.1 |
| 13 | 364 | 788 | 53.8 |
| 14 | 274 | 538 | 49.1 |
| 15 | 188 | 346 | 45.7 |
| 16 | 200 | 358 | 44.1 |
| 17 | 286 | 550 | 48.0 |
| 18 | 360 | 730 | 50.7 |
| 19 | 290 | 608 | 52.3 |
| 20 | 376 | 800 | 53.0 |
| 21 | 450 | 980 | 54.1 |
| 22 | 364 | 788 | 53.8 |
| 23 | 286 | 550 | 48.0 |
| 24 | 372 | 742 | 49.9 |
| 25 | 446 | 922 | 51.6 |
| 26 | 376 | 800 | 53.0 |
| 27 | 462 | 992 | 53.4 |
| 28 | 536 | 1,172 | 54.3 |
| 29 | 450 | 980 | 54.1 |
| 30 | 360 | 730 | 50.7 |
| 31 | 274 | 538 | 49.1 |
| **Total** | **10,810** | **21,243** | **49.1** |

The modular approach significantly reduces development effort, allowing for scenario reusability, better readability (with the main scenario file primarily containing customizable elements, while the scenario implementation remains within simple scenario files), and easy customization through the use of parameter values.

It is essential to note that the repeated use of scenarios within a modular scenario leads to a much more optimized codebase compared to a traditional single-file scenario.

## 5 Conclusion

Modular2Simple effectively addresses the challenge of complex scenario creation in ADS by enabling the construction and reuse of modular scenarios, leading to more efficient and realistic testing. By utilizing the OpenSCENARIO format and integrating seamlessly with simulators like CARLA, Modular2Simple enables developers to create and manage modular scenarios through a straightforward command-line interface. The tool supports the creation and handling of modular scenarios by allowing the combination of multiple simple or modular scenarios into more complex scenarios. This approach not only promotes scenario customization and reuse but also provides greater flexibility and reduces redundancy and effort associated with scenario creation. Modular2Simple advances ADS scenario development by enhancing modularity, facilitating reuse, and supporting flexible scenario management, making it a valuable tool for the ADS research and development community.

## 6 Limitations and Future Research

Despite its contributions, the proposed method has certain limitations. The primary limitation lies in the compatibility between the OpenSCENARIO format and simulation environments such as CARLA. Currently, not all OpenSCENARIO tags and functionalities are fully translated or supported within CARLA, leading to potential discrepancies in scenario representation. The same issue may occur with other simulation environments that have varying levels of OpenSCENARIO support. Additionally, the lack of a robust predefined module library limits the tool's flexibility. Developing an extensive library with different possible actors and scenarios that ADS might encounter could add significant value, enabling researchers to design complex, realistic scenarios more efficiently.

To overcome this limitation and further improve Modular2Simple, several directions for future work are suggested. One potential area is to extend the tool's compatibility with additional scenario formats, making it more widely usable across different simulation environments. Another area for development is improving the translation and mapping of unsupported OpenSCENARIO elements to ensure consistent scenario behavior across platforms. Additionally, creating a comprehensive module library with predefined actors and scenarios could facilitate rapid development and enhance the tool's utility for the ADS community. Exploring these future directions would help establish Modular2Simple as a more robust and versatile tool for ADS scenario creation, increasing its value for researchers and developers in the field of autonomous driving.

**Acknowledgments**

This study is part of the CAS Project for young Scientists in Basic Research (YSBR-40) and ISCAS New cultivation Project (ISCAS-PYFX-202201). We are grateful to the National Natural Science Foundation of China (62172217) and the Chinese Academy of Sciences for funding this work.

**References**

1. Dosovitsky A., Ros G., Cadevilla F., Lopez A. and Koltun V. CARLA: An Open Urban Driving Simulator. In Proceedings of the 1st Annual Conference on Robot Learning, pages 1-16, Mountain View, CA, USA.
2. ASAM OpenSCENARIO: User Guide. On-line: https://releases.asam.net/OpenSCENARIO/1.0.0/ASAM_OpenSCENARIO_BS-1-2_User-Guide_V1-0-0. html, Accessed: 20/11/2023.
3. https://github.com/NikolaiKhriapov/modular2simple
4. Kalra N. and Paddock S. M. Driving to safety: How many miles of driving would it take to demonstrate autonomous vehicle reliability? Transportation Research Part A: Policy and Practice, 94:182– 193, 2016.
5. Queiroz R., Berger T., and Czarnecki K. Geoscenario: An open dsl for autonomous driving scenario representation. In Proceedings of 2019 IEEE Intelligent Vehicles Symposium (IV), pages 287–294, Paris, France, 2019.
6. Bock F., Sippl C., Heinzz A., Lauerz C. and German R. Advantageous usage of textual domain-specific languages for scenariodriven development of automated driving functions. In Proceedings of the 2019 IEEE International Systems Conference (SysCon), pages 1-8, Orlando, FL, USA, 2019.
7. Schütt B., Braun T., Otten S. and Sax E. SceML: A graphical modeling framework for scenario-based testing of autonomous vehicles. In Proceedings of the 23rd ACM/IEEE International Conference on Model Driven Engineering Languages and Systems, pages 114-120, Virtual, 2020.
8. Goss Q., AlRashidi Y. and Akbaş M. I. Generation of Modular and Measurable Validation Scenarios for Autonomous Vehicles Using Accident Data. In Proceedings of the 2021 IEEE Intelligent Vehicles Symposium (IV), page 251-257, Nagoya, Japan, 2021.
9. Ahmed A., Sadullah A.F.M. and Yahya, A.S. Errors in Accident Data, Its Types, Causes and Methods of Rectification-Analysis of the Literature. Accid. Anal. Prev. 130:3–21, 2019.
10. Otte D., Jänsch M. and Haasper C. Injury Protection and Accident Causation Parameters for Vulnerable Road Users Based on German In-Depth Accident Study GIDAS. Accid. Anal. Prev., 44:149–153, 2012.
11. Nalic D., Eichberger A., Hanzl., Fellendorf M. and Rogic B. Development of a Co-Simulation Framework for Systematic Generation of Scenarios for Testing and Validation of Automated Driving Systems. In Proceeedings of IEEE Intelligent Transportation Systems Conference (ITSC), page 1895-1901, Auckland, New Zealand, 2019.
12. Waymo L, On the road to fully self-driving. On-line: https://waymo.com/safety/, Accessed: 20/11/2023.